\documentstyle[12pt,aaspp4,psfig]{article}
\begin{document}
\newcommand{\BB}{\mbox{${\bf B}_0$}}
\newcommand{\beq}{\begin{equation}}
\newcommand{\eeq}{\end{equation}}
\newcommand{\lsim}{\mbox{$\stackrel{<}{\scriptstyle\sim}$}}
\newcommand{\gsim}{\mbox{$\stackrel{>}{\scriptstyle\sim}$}}
\newcommand{\D}[2]{\makebox{$\displaystyle\frac{\partial{#1}}{\partial{#2}}$}}
\newcommand{\DD}[2]{\makebox{$\displaystyle\frac{\partial^2{#1}}{\partial{#2}^2}$}}
\title{Stochastic Acceleration of Low Energy Electrons in Cold Plasmas.}
\author{Julia M. Pryadko\altaffilmark{1} and Vah\'{e} Petrosian\altaffilmark{2}}
\affil{Center for Space Science and Astrophysics\\
Stanford University\\
Stanford, CA 94305-4055}
\date{today}

\altaffiltext{1}{Department of Physics}
\altaffiltext{2}{Department of Physics and Applied Physics}


\begin{abstract}
  We investigate the possibility of stochastic acceleration of
  background low-energy electrons by turbulent plasma waves.  We
  consider the resonant interaction of the charged particles with all
  branches of the transverse plasma waves propagating parallel to a
  uniform magnetic field. Numerical results and asymptotic analytic
  solutions valid at non-relativistic and ultra-relativistic energies
  are obtained for the acceleration and scattering times of electrons.
  These times have a strong dependence on plasma parameter $\alpha =
  \omega_{pe}/\Omega_e$ (the ratio of electron plasma frequency to
  electron gyrofrequency) and on the spectral index of plasma
  turbulence.  It is shown that particles with energies above certain
  critical value may interact with higher frequency electromagnetic
  plasma waves and this interaction is allowed only in plasmas with
  $\alpha <1$. We show that for non-relativistic and semi-relativistic
  electrons in low-$\alpha$ plasmas the ratio of the acceleration time
  to the scattering time can be less than unity for a wide range of
  energies. From this we conclude that the transport equation derived
  for cosmic rays which requires this ratio to be much larger than one
  is not applicable at these energies.  An approximate "critical"
  value of particle energy above which the dynamics of charged
  particles may be described by this transport equation is determined
  as a function of plasma parameters.  We propose new transport
  equation for the opposite limit (energies less than this critical
  value) when the acceleration rate is much faster than the pitch
  angle scattering rate.  This equation is needed to describe the
  electron dynamics in plasmas with $\alpha \lesssim 0.1$.

\end{abstract}

\section{INTRODUCTION}
The acceleration and propagation of charged particles in magnetized
plasmas via their stochastic
interaction with turbulent plasma waves is a problem of wide interest 
in different astrophysical areas. 
The interaction of the particles with plasma turbulence causes 
diffusion of the energetic particles in phase space and
leads to stochastic
acceleration which is a second order Fermi process.
The rates of these
interactions are controlled by the magnetic field
$B_0$ and other physical properties of the background
plasma such as the
background particle density $n$, the energy density and spectrum of
the turbulence and, if in thermal equilibrium, the background 
temperature $T$. 

The influence of different modes of plasma waves on the dynamics of 
high energy charged particles has been
 discussed in the literature for different plasma conditions.  
 Whistler
waves  for electrons in an electron-proton background
plasma for solar flare conditions have been considered by Steinacker 
and Miller (1992), Miller (1991), Miller and Ramaty (1987),
Hamilton and Petrosian (1992)
and by Achatz et al (1993) for interplanetary conditions. 
Benz (1977) and Melrose (1980) consider Langmuir (plasma) waves,
Achterberg (1981) and Beaujardiere and Zweiles (1989) deal with
magnetosonic waves, Benz and Smith (1987) describe interaction with lower
hybrid waves.
The transport equation for transverse Alfv\'en
waves in a cold background medium has been considered by Schlickeiser (1989)
and  this consideration have been extended for electromagnetic branch in general 
astrophysical plasmas by Dung and Petrosian (1994) (hereafter 
refered to as DP).
Most of these and other works investigating particle acceleration deal
with relativistic particles and assume an initial source 
of non-thermal high energy particles. 

We are interested in the investigation of acceleration of the particles
present in the background. These particles usually have a Maxwellian
distribution with $kT \ll mc^2$ and therefore have low energies.
The general problem is formulated in DP and some results are given for
high and intermediate energies.
In this paper we continue this investigation of interaction with
all modes of transverse plasma waves propagating parallel and/or 
antiparallel to
the ordered magnetic field with emphasize on 
{\em low energy electrons}.
In \S2 we give an overview of the basic equations
of turbulent plasma theory used in this paper and define a ratio of scattering 
to acceleration times in terms of  Fokker-Planck coefficients which are
discussed in \S3.
In \S4 we evaluate this ratio for different plasmas
and particles of different energies and suggest a new transport
scenario for low energy particles.
In \S5 and \S6 we derive approximate analytic expressions for the above
ratio and the acceleration time for non-relativistic and extremely
relativistic electrons, respectively.
A brief summary and our conclusions are presented in \S7.

\section{BASIC EQUATIONS}
We consider the behavior of the
energetic charged particles in a background plasma with uniform
magnetic field of strength $B$ and superposed
plasma waves. The gyrophase-averaged phase space density $f(z,t,p,\mu)$
then obeys the differential Fokker-Planck equation (Schlickeiser 1989):
\beq\label{f-p}
        \D{f}{t}+v\mu\D{f}{z}=
        \frac{1}{p^2}\D{}{p}p^2\left (D_{pp}\D{}{p}+D_{p \mu}\D{}{\mu}
        \right ) f
        +\D{}{\mu}\left ( D_{\mu\mu}\D{}{\mu}+D_{\mu p}\D{}{p} \right )f+S.
\eeq

Here $z$ is the distance along the field lines, $S$ is the source
function, $\mu$ denotes the cosine of the pitch angle, $v=
p/(m\gamma)$ is the velocity and $\gamma = (1+p^2/(mc)^2)^{1/2}$ is 
the Lorentz factor of the particle with momentum $p$. 
The interaction of the particles with plasma waves is
described by the three Fokker-Plank coefficients $D_{\mu \mu}, D_{\mu
  p}=D_{p \mu}$ 
and $D_{pp}$ which
depend on the properties of the turbulence.

Solving the differential equation (\ref{f-p}) in general is not simple
so one usually considers the solution for certain physical
conditions when it can be simplified.
In cases when the pitch angle scattering time scale
$\tau_{sc}(\mu)\simeq D_{\mu\mu}^{-1}$ is much shorter than 
the traverse time $\tau_{tr}\simeq L/v$, where $L$ is the size of the
turbulent plasma region, 
the pitch angle distribution will be nearly isotropic
so that the anisotropic
part of the phase space density $g(z,t,p,\mu)=f(z,t,p,\mu)-F(z,t,p)$
is much smaller than the pitch angle averaged phase
space density
\beq
        F(z,t,p)=\frac{1}{2}\int_{-1}^{+1}d\mu f(z,t,p,\mu).
\eeq
In addition if the ratios
\beq\label{r}
R_1(\mu, p)=\frac{D_{pp}/p^2}{D_{\mu\mu}}\ \ll 1,\ \ \ 
R_2(\mu, p)=\frac{D_{p \mu}/p}{D_{\mu\mu}}\ \ll 1
\eeq
then equation (\ref{f-p}) can be reduced to the 
diffusion-convection
equation which is also known as the transport equation for the cosmic rays
(Jokipii 1966; Kirk et al. 1988; Schlickeiser 1989; DP):
\beq\label{trans}
        \D{F}{t}=\D{}{z}\kappa_1\D{F}{z}+(p v)\D{\kappa_2}{z}\D{F}{p}-
        \frac{1}{p^2}\D{}{p}(p^3 v\kappa_2)\D{F}{z}+
        \frac{1}{p^2}\D{}{p}(p^4\kappa_3\D{F}{p})+Q(z,t,p).
\eeq
Here $Q(z,t,p)$ is the pitch-angle averaged source term and the three
transport coefficients are defined as
\beq\label{kappa1}
\kappa_1=\frac{v^2}{8}\int_{-1}^{+1}d\mu\frac{(1-\mu^2)^2}{D_{\mu\mu}}
\eeq
\beq\label{kappa2}
\kappa_2=\frac{1}{4}\int_{-1}^{+1}d\mu(1-\mu^2) R_2
\eeq
\beq\label{kappa3}
\kappa_3=\frac{1}{2}\int_{-1}^{+1}d\mu D_{\mu\mu} (R_1-R_2^2)
\eeq
The first four terms of the right hand side of equation (\ref{trans}) represent
 spatial diffusion, adiabatic
acceleration/deceleration, spatial convection and momentum diffusion.
If the above equations are satisfied we can define averaged
scattering, acceleration and spatial diffusion times as $\tau_{sc}=
8 \kappa_1/v^2$, 
$\tau_{ac}=1/\kappa_3$, $\tau_{diff}=L^2/\kappa_1=8\tau_{tr}^2/\tau_{sc}$,
respectively.

The assumptions that lead to transport equation (\ref{trans}) are not
always valid. While the condition $\tau_{sc}
\ll L/v$ holds for wide range of particle energy in most astrophysical 
plasmas the requirements in equation (\ref{r}) are not always
satisfied. As shown below these ratios can exceed unity for low
energy background particles.

The goal of this paper is to investigate the range of validity of
equation (\ref{trans}) 
and suggest a different transport scenario for cases when this equation 
is inapplicable. Here we will deal only with waves propagating
parallel or antiparallel to the magnetic field. The influence of transverse
plasma waves propagating perpendicular to the magnetic field on the dynamics
of charged particles will be considered in a subsequent paper.

\section{FOKKER-PLANCK COEFFICIENTS}

Following DP (see also Lerche 1968) 
and assuming a power law distribution of plasma turbulent energy density
as a function of wave vector,
${\cal E}(k)=(q-1){\cal E}_{tot}\ K_{min}^{q-1}\ K^{-q}$
(for $K \geq K_{min}$ and $q>1$), it can be shown that
the Fokker-Planck coefficients necessary for evaluation of the ratios
in equation (\ref{r}) can be written as:
\beq\label{dmm}
D_{\mu\mu}=\frac{1}{\tau_p\gamma^2}(1-\mu^2)\sum_{j=1}^{N}
\left( 1-\mu\frac{\beta_{ph}(k_j)}{\beta}\right)^2\chi(k_j)
\eeq
\beq
\frac{D_{\mu p}}{p}=\frac{1}{\tau_p\gamma^2}(1-\mu^2)\sum_{j=1}^{N}
\frac{\beta_{ph}(k_j)}{\beta}
\left( 1-\mu\frac{\beta_{ph}(k_j)}{\beta}\right)\chi(k_j)
\eeq
\beq\label{dpp}
\frac{D_{pp}}{p^2}=\frac{1}{\tau_p \gamma^2}(1-\mu^2)\sum_{j=1}^{N}
\left(\frac{\beta_{ph}(k_j)}{\beta}\right)^2\chi(k_j),
\eeq
where
\beq
\chi(k_j)=\frac{|k_j|^{-q}}{|\beta\mu - \beta_{gr}(k_j)|},
\eeq
$\beta = v/c$ and the dimensionless wave vector $k_j$ 
is one of the roots (maximum of four) of the resonant condition:
\beq\label{res}
\omega(k_j)-\mu\beta k_j \mp 1/\gamma =0, \ \ \ \ k_j = K_jc/\Omega_e. 
\eeq
Here and in what follows the upper and lower signs refer to 
the right(R) and left(L) hand polarized plasma modes.
The wave frequency $\Omega$, in units of
electron gyrofrequency $\Omega_e$, 
is determined from the dispersion relation:
\beq\label{disp}
\frac{k^2}{\omega^2}= 
1-\frac{\alpha^2 (1+\delta)}{(\omega \mp 1)(\omega \pm \delta)}, \ \ \
\  \omega(k_j)=\Omega(k_j)/\Omega_e,
\eeq
where 
 $\delta = \frac{m_e}{m_i}$
 is the ratio of electron to proton masses and
\beq\label{alpha}
\alpha =\omega_{pe}/\Omega_e = 3.2\ (n_e/10^{10}\ {\rm cm}^{-3})^{1/2}
(B /100\ {\rm G})^{-1}
\eeq 
is the ratio of electron plasma frequency to
gyrofrequency which is simply related to the Alfv\'en
velocity $\beta_a$ expressed in units of speed of light as
\beq\label{alvel}
\beta_a=\frac{\sqrt{\delta}}{\alpha}.
\eeq 
The phase and group velocities of these
waves (in units of speed of light): $\beta_{ph}(k_j) = \omega_j/k_j$ and
$\beta_{gr}(k_j) = d \omega_j/d k_j$, respectively, can be
obtained from relations (\ref{res}) and (\ref{disp}). 
The parameter $\tau_p$,
which is a typical time scale in the turbulent plasma, is defined as 
\beq\label{taup}
\tau_p^{-1}=\frac{\pi}{2}\Omega_e\left(\frac{{\cal E}_{tot}}{B^2/(8\pi)}\right)
(q-1)k_{min}^{q-1}.
\eeq
The parameters important for our problem are $\Omega_e$, $\alpha$, $q, k_{min}$
and the ratio of plasma turbulent density to magnetic energy density;
$f_{turb}=(8\pi{\cal E}_{tot}/B^2)$. The above equations hold for
both electrons and protons. In what follows we consider only interaction
and acceleration of electrons.

\subsection{Critical Angles}

As described in DP, in general,
four values of $k_j$ contributes to
the Fokker-Planck coefficients, except for $\mu=0$ and $\gamma<\delta^{-1}$
and for the pitch angles between
the two critical values when only two values of $k_j$ are
involved. For non-relativistic electrons one of these roots is
due to resonant interaction with ion-cyclotron waves with $k_j \gg 1$,
so that its
contribution to the coefficients is negligible ($\chi(k_j) \ll 1$).
The main
contribution to the Fokker-Planck coefficients then comes from the 
electron's interaction with 
whistler mode and higher-frequency electromagnetic branch. 
The last interaction, allowed only 
in low $\alpha$-plasmas ($\alpha <1$), occurs for 
pitch angles greater than some critical angle $\mu_{cr}^{em}$ at which 
the group velocity of
electromagnetic wave becomes equal to the component of the electron
velocity along the magnetic field, $\beta_{gr} = \mu \beta$.
Similarly,
there exists another critical angle $\mu_{cr}^{ec} < \mu_{cr}^{em}$ 
for interaction of electrons with
electron-cyclotron waves (the high-$k$ end of the branch which is
commonly refered to as
whistler branch at $\omega(k) \ll 1$). 
It is well known that for these angles the quasilinear approximation breaks
down and the Fokker-Planck coefficients become infinite. 
This has a minor consequence for the general process considered here
(Steinacker and Miller 1992, DP). However, in order to clarify some of the 
behaviors of the coefficients we give a brief description of these angles.

\begin{figure}[tbp]
  \psfig{file=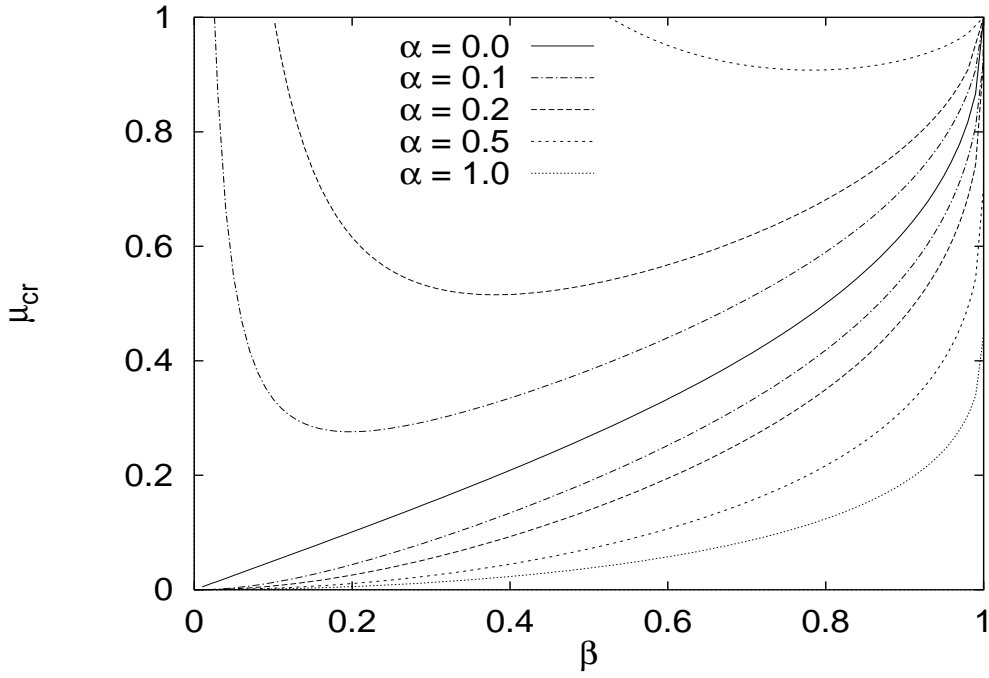,width=5.5in,height=3.5in,angle=0}
\caption{
The dependence of the critical angles $\mu_{cr}^{ec}$ (lower
curves) and $\mu_{cr}^{em}$ (upper curves) on electron's velocity 
$\beta$ for different plasma parameters $\alpha$. For $\alpha = 0$
(the empty space) $\mu_{cr}^{ec} = \mu_{cr}^{em}$ (solid line).}
\label{mucr}
\end{figure}

The dependence of these two critical angles on electron velocity
$\beta$ and plasma parameter $\alpha$ are shown on Figure \ref{mucr}.
The upper and lower curves correspond to $\mu_{cr}^{em}$ and
$\mu_{cr}^{ec}$ respectively. 
Thus for $0<\mu< \mu_{cr}^{ec}$ there are three roots
coming from the interaction with forward (2 roots) and backward (1
root) moving
electron-cyclotron waves.  Similarly, for $1 > \mu > \mu_{cr}^{em}$ 
three are roots, two from the electromagnetic branch and one 
from the backward
moving electron-cyclotron branch. But for $\mu_{cr}^{em} > \mu > \mu_{cr}^{ec}$
only the later root exists.
The fourth and unimportant root from
the (backward) ion-cyclotron branch mentioned above is common to
all three cases.
Same situation holds for negative values of $\mu$ with the role of the
forward and backward waves reversed.

\begin{figure}[tbp]
  \psfig{file=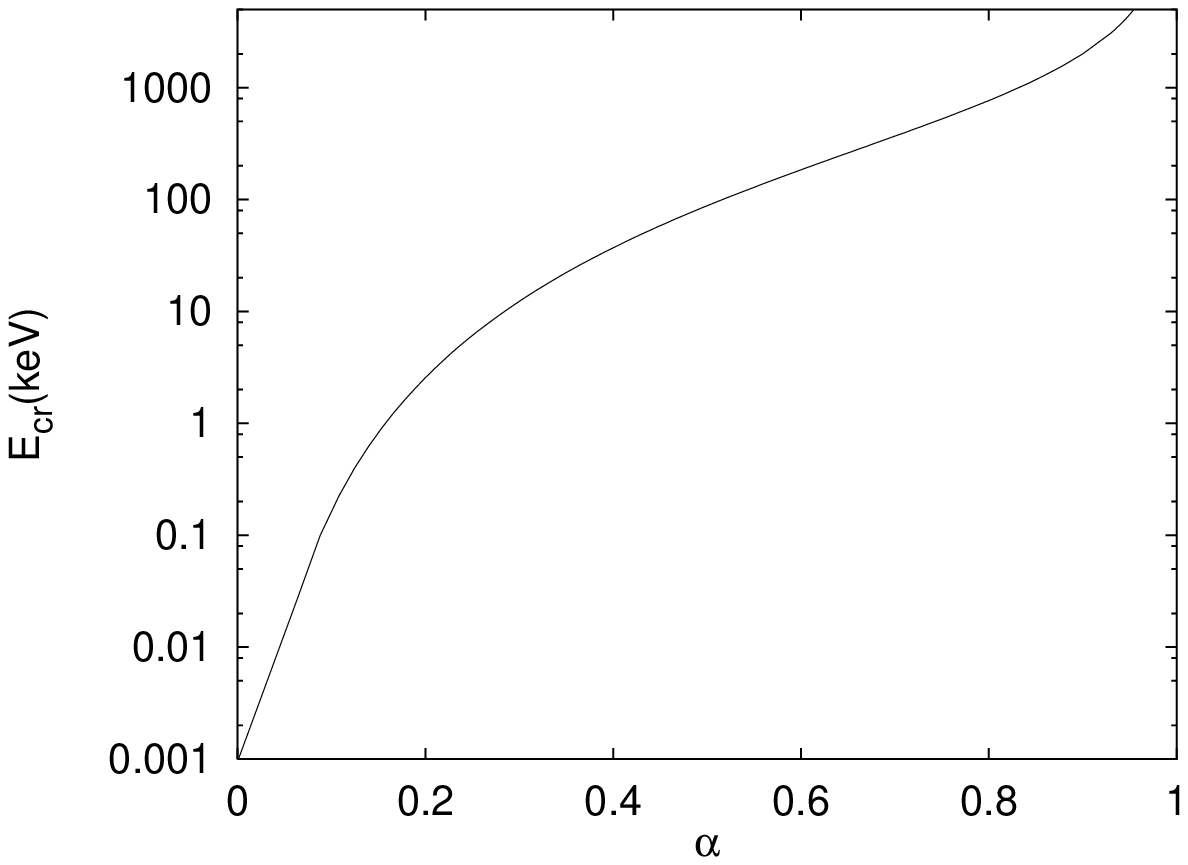,width=5.5in,height=3.5in,angle=0}
\caption{
  The plasma parameter dependence of the critical value of
  electron kinetic energy, below which interaction with the upper
  electromagnetic branch is not allowed.}
\label{goft}
\end{figure}

For each value of
$\alpha$ there exist a critical velocity $\beta_{cr}$ or kinetic
energy $E_{cr}$ below which $\mu_{cr}^{em} = 1$ and 
interaction of electrons with electromagnetic branch is not allowed.
Figure \ref{goft}
shows the variation of $E_{cr}$ with the
plasma parameter $\alpha$.
For high values of $\alpha$ (high plasma density, low magnetic field)
$\beta_{cr} \rightarrow 1$.
In the opposite case of $\alpha \rightarrow 0$ (
very high magnetic field and/or low plasma density) the two curves merge
to the solid line on Figure \ref{mucr} which is described by 
the expression $\mu_{cr} = (1 - \sqrt{1-\beta^2})/\beta$.

\begin{figure}[tbp]
  \psfig{file=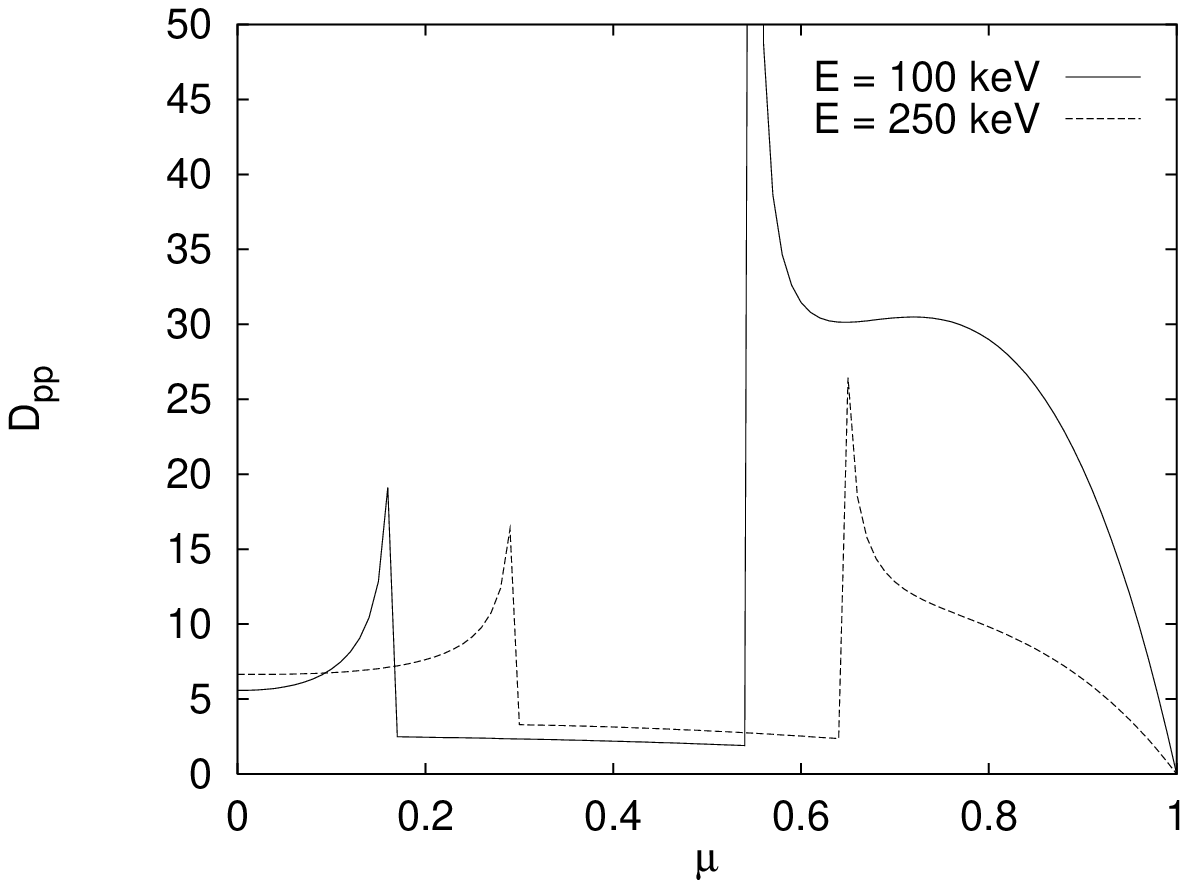,width=5.5in,height=3.1in,angle=0}
\caption{
  The dependence of the Fokker-Planck coefficient $D_{pp}$ on pitch
  angle for electrons with two different values of kinetic energy in
  plasma with $\alpha = 0.2$ and $q=1.6$. For smaller values of
  $\alpha$ the flat, single root regime in the middle range of $\mu$
  becomes smaller. This regime extends to the whole range of $\mu$ for
  larger values of $\alpha$ and non-relativistic energies.}
\label{dppm}
\end{figure}

The effect of this behavior can be seen on Figure \ref{dppm} 
where we show the dependence of one of the Fokker-Planck
coefficients, $D_{pp}$, on electron pitch angle in a plasma with
$\alpha = 0.2$.
As we approach one of the critical values of
the pitch angle $\mu_{cr}^{em}$ or $\mu_{cr}^{ec}$ the coefficient 
becomes infinite as indicated by the sharp cusps. For each energy the
first and second peaks occur at $\mu_{cr}^{ec}$ and $\mu_{cr}^{em}$, 
respectively.

\section{RATIOS $R_1$ AND $R_2$.}

Using equations (\ref{dmm}) to (\ref{dpp}) the ratios
defined in equation (\ref{r}) become 
\begin{eqnarray}\label{ratio}
R_1&=&
\frac{\sum_{j=1}^{N}
\beta_{ph}(k_j)^2\chi(k_j)}
{\sum_{j=1}^{N}
\left( \beta-\mu \beta_{ph}(k_j)\right)^2\chi(k_j)},\nonumber \\
R_2&=&
\frac{\sum_{j=1}^{N}
\beta_{ph}(k_j)\left(\beta-\mu \beta_{ph}(k_j)\right)\chi(k_j)}
{\sum_{j=1}^{N}
\left( \beta-\mu \beta_{ph}(k_j)\right)^2\chi(k_j)},
\end{eqnarray}
which depend on the electron pitch angle and velocity, and on the 
plasma parameters $\alpha$ and $q$. Often, in particular for 
$\mu_{cr}^{em}> \mu >\mu_{cr}^{ec}$, only one root has significant 
contribution. In this case the above expressions simplify to
$R_1=R_2^2= \omega^2/(\beta k - \mu \omega)^{-2}$.

Since the
derivation of transport equation (\ref{trans}) requires these ratios
 to be much less
than 1 we want to determine the plasma conditions and energy range for
which this requirement is satisfied. It is well known that
relativistic electrons (and all protons),
 interact mainly with Alfv\'en waves. In this case (see also \S 6 below)
$R_1 \propto (\beta_a/\beta)^2 \ll 1$ if
$\beta_a \ll 1$ for relativistic electrons (or $\beta \ll \beta_a$ for
protons). 
This condition tends to breakdown as we
go to the lower
energies, where the interaction of the electrons with other plasma
waves become more significant than the interaction with Alfv\'en and
ion-cyclotron waves.

\subsection{Electrons with $90 \arcdeg$ Pitch Angles.}

\begin{figure}[tbp]
  \psfig{file=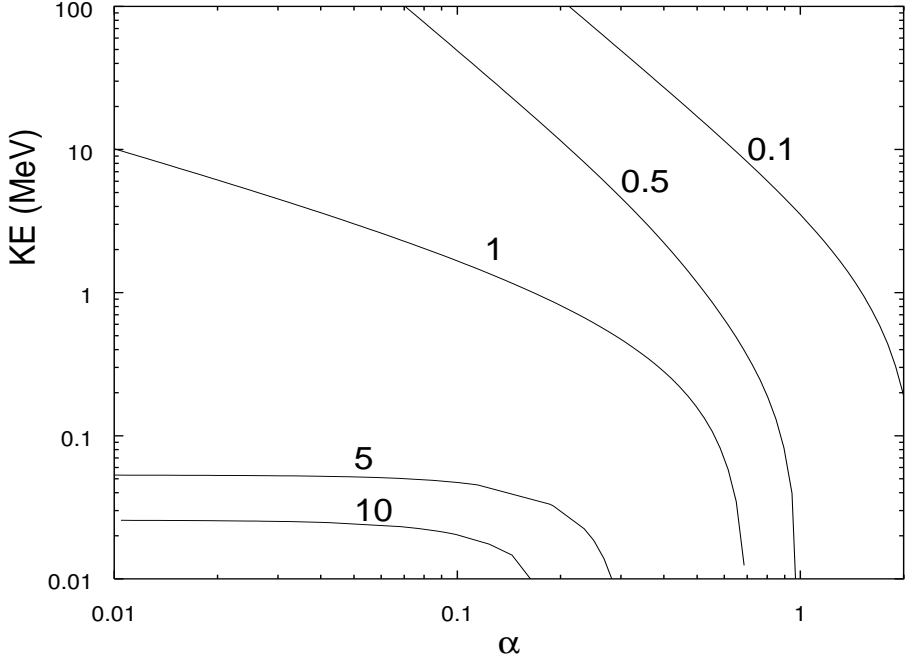,width=5.5in,height=3.5in,angle=0}
\caption{
  The dependence of the electron kinetic energy $E$ on plasma
  parameter $\alpha$ for five different ratios $R_1$
  for electrons with pitch angle $\mu = 0$.}
\label{ratio1}
\end{figure}

The above equations can be solved analyticly for $\mu = 0$ when the
details of the distribution ${\cal E}(k)$ are not important. For
this case  the resonant condition (\ref{res})
is simplified to $\omega(k_j)=\pm 1/\gamma$ and the
ratio $R_1=\beta_{ph}^2(k)/\beta^2$. The ratio $R_2 = 0$ because
$D_{\mu p } = 0$.
Substitution of this in equation (\ref{disp}) gives 
four symmetric roots (two from the Alfv\'en branch and two from the 
whistler branch)
which differ only in their signs
of $\omega_j$ or $k_j$. We then have the analytic relation
\beq\label{rzeromu}
R_1=\frac{\gamma^2 (1 + \gamma \delta)}{(1+\gamma)(\gamma^2
  \alpha^2+\gamma-1)}, \ \ \ \ R_2 = 0.
\eeq
In the extreme relativistic regime $\gamma \gg \delta^{-1}$, $R_1 =
\frac{\delta}{\alpha^2 + \delta} < 1$ and for $1 \ll \gamma \ll
\delta^{-1}$, $R_1 = \gamma^{-1} ( \alpha^2 + \delta)^{-1} \ll 1$, but
for non-relativistic electrons $R_1 \simeq 1/2 \alpha^2$ which can
exceed unity if $\alpha^2 < 1/2$.
Figure \ref{ratio1} shows the relation between the electron kinetic
energy and the plasma parameter for five different values of this ratio.
As evident, for small $\alpha$ the condition 
(\ref{r}) required for validity of equation (\ref{trans}) is
violated up to very high energies (Note that for
$\alpha<<\sqrt{\delta} \simeq 0.024$ this is true at all energies).
For example, for plasmas with $\alpha \lesssim 0.1$ 
the diffusion approximation becomes valid
only for electrons with kinetic energy exceeding few MeV. 
This means that for low energy electrons in low-$\alpha$ plasmas we
cannot use equation
(\ref{trans}) and we must revert back to the original Fokker-Planck
equation (\ref{f-p}). 

In the opposite case of $R_1 \gg 1$ (low energies, $\alpha^2 \ll 1$) 
the scattering time becomes much
greater than the acceleration time so that the electron pitch angle 
will
not change significantly in times of order of the acceleration 
time scale. We can, therefore, neglect the small  
$D_{\mu \mu}$ and $D_{\mu p} = 0$   
terms in equation (\ref{f-p}) and obtain
the simple diffusion equation
\beq\label{newtr0}
\D{f}{t}=
\frac{1}{p^2}\D{}{p}p^2 D_{pp} \D{f}{p} + S,\ \ \ \mu=0.        
\eeq
Thus for this special case of $\mu=0$ (i.e. $90^o$ pitch angle) 
the Fokker-Planck equation 
reduces to the pure diffusion equation
in momentum (or energy) space.
>From inspection of Figure \ref{ratio1} we can see that this equation must be
used up to several hundreds of keV for $\alpha \leq 0.3$.

\subsection{Electrons with $\mu \neq 0$}

\begin{figure}[tbp]
  \psfig{file=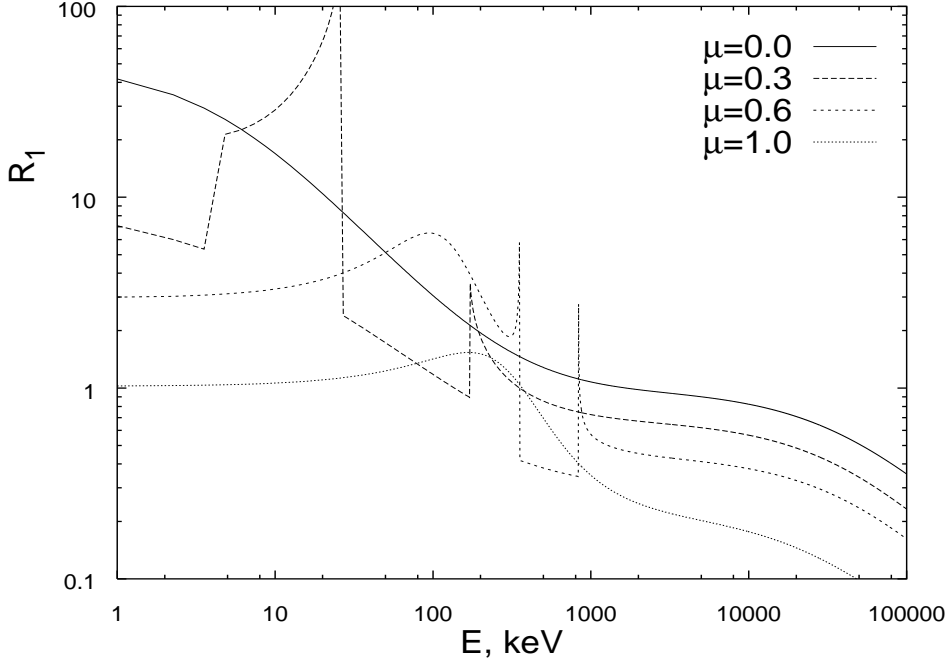,width=5.5in,height=3.5in,angle=0}
\caption{
  The dependence of the ratio $R_1$ on electron energy for four
  different values of pitch angle in the plasma with parameter
  $\alpha=0.1$ and spectral index $q=1.6$.}
\label{ratmu}
\end{figure}

\begin{figure}[tbp]
  \psfig{file=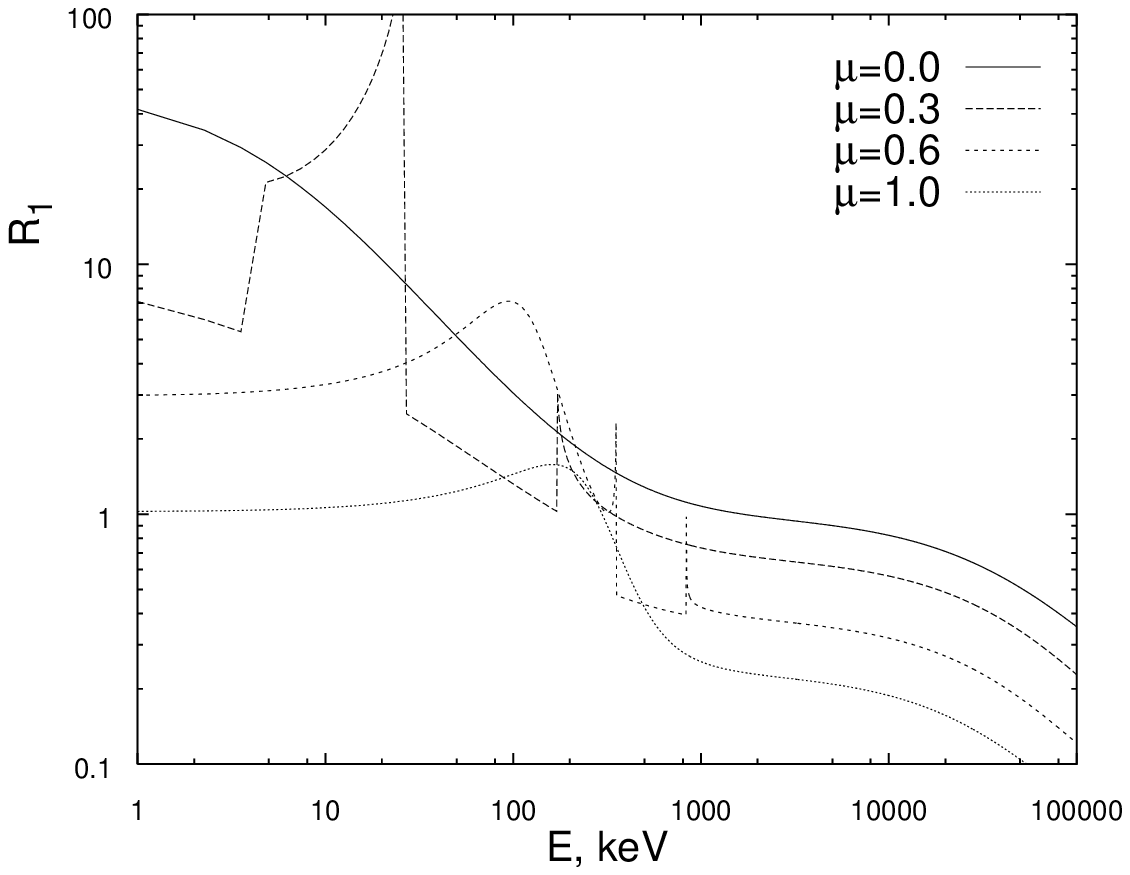,width=5.5in,height=3.5in,angle=0}
\caption{
  Same as Figure \protect\ref{ratmu} with $\alpha=0.1$ and $q=3$.}
\label{ratmu013}
\end{figure}

\begin{figure}[tbp]
  \psfig{file=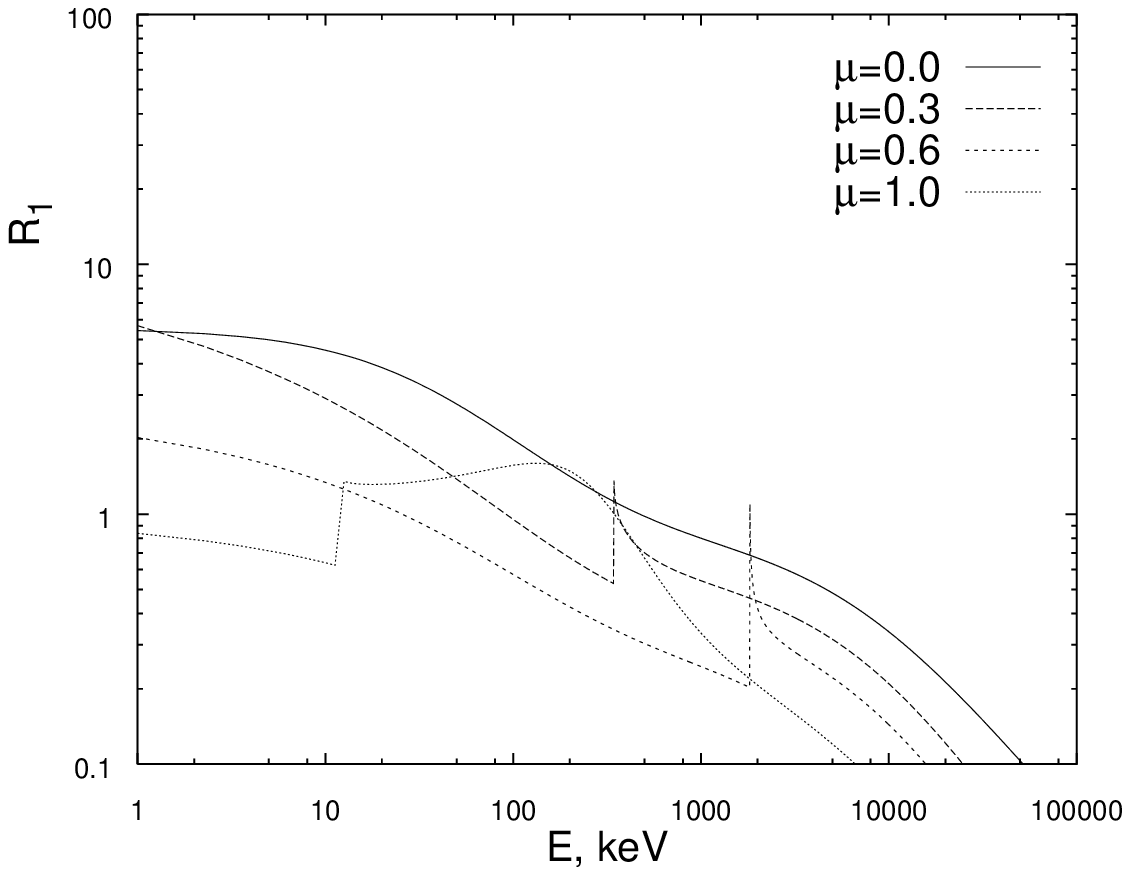,width=5.5in,height=3.5in,angle=0}
\caption{
  Same as Figure \protect\ref{ratmu} with $\alpha=0.3$ and $q=1.6$.}
\label{ratmu03}
\end{figure}

\begin{figure}[tbp]
  \psfig{file=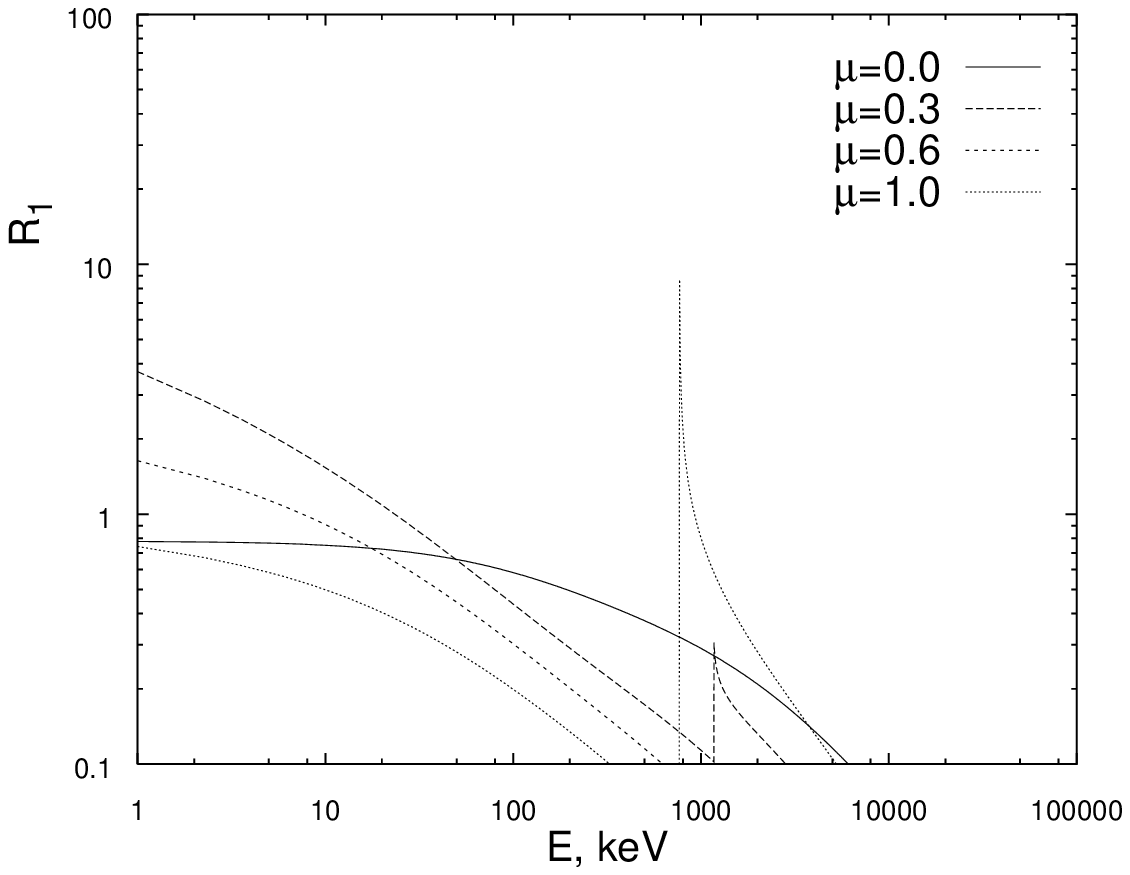,width=5.5in,height=3.5in,angle=0}
\caption{
  Same as Figure \protect\ref{ratmu} with $\alpha=0.8$ and $q=1.6$.}
\label{ratmu08}
\end{figure}

The situation is more complicated for electrons with $\mu \neq 0$
because the resonant condition (\ref{res}) is no longer simple and
there are multiple roots $k_j$ contributing in the summations in
equations (\ref{ratio}). Analytic approximation for the
ratios is not possible in this case and we have to evaluate these ratios
numerically. Figures \ref{ratmu} - \ref{ratmu08} show variation of $R_1$
with energy for different values of $\mu$ in plasmas with four
different values of $\alpha$. As evident, these ratios
vary considerably with $\mu$ but in general they tend to increase with
decreasing energy. Furthermore, they depend on the plasma parameter
$\alpha$ and the spectral index $q$. The ratios
$R_1$ and $R_2$ tend to increase with decreasing $\alpha$ (high
magnetic field, low plasma density) but they are not very
sensitive to $q$ (compare Figures \ref{ratmu} and \ref{ratmu013}). 
The discontinuous behavior of the curves arise from the
existence of the two critical pitch angles discussed in
the \S 3.1.
For given $\alpha$ and $\mu$ there exists a maximum of three and
a minimum of one energy for which this value of $\mu$
is critical and leads to a discontinuity (except $\mu = 0$, 
which is not critical for any energy).  
Unlike the Fokker-Planck coefficients,
their ratios do not become infinite at critical values of
$\mu$ or energy.
In \S 5 we will derive the
asymptotic expression for the ratios $R_1$ and $R_2$ as a function of
$\alpha$ for low energy electrons. 

Following the same argument used in the case of zero pitch angle
electrons we can conclude that for low energy electrons 
the distribution function
can be obtained from the solution of the following equation
\beq\label{newtr}
\D{f^{\mu}}{t}+v\mu\D{f^{\mu}}{z}=
\frac{1}{p^2}\D{}{p}p^2 D_{pp}^{\mu} \D{f^{\mu}}{p} + S^{\mu},        
\eeq
where $D_{pp}^{\mu}$ is given by equation (\ref{dpp}) for each $\mu$.
The advective term (second on the left hand side) can also be
neglected if the traverse time $L/v$ is much larger than the
acceleration time $p^2/D_{pp}$ and if $D_{pp}$
and the source term $S^{\mu}$ are independent of $z$
(homogeneous plasma). Equation (\ref{newtr}) then becomes identical to
equation (\ref{newtr0}) for zero-pitch angle electrons. 
Once the electron achieve higher energies and moves above the 
line $R_1 = 1$ so that the conditions (\ref{r}) are satisfied, then 
we revert to equation (\ref{trans}).
In the transition region neither equation (\ref{trans}) nor
(\ref{newtr}) are valid and the resulting 
equation for isotropic pitch angle distribution 
is almost as complicated as the original equation (\ref{f-p}).

\section{NON-RELATIVISTIC APPROXIMATIONS}

For low energy electrons $\gamma \rightarrow 1+\beta^2/2 $ and the
resonance condition (\ref{res}) combined with dispersion relation
(\ref{disp}) gives
\beq\label{cubeq} \mu\beta k^3 - (\frac{1}{2} + 2 \mu^2)\beta^2 k^2 -
\mu \beta (1 -\alpha^2) k +\alpha^2=0,
\eeq
where we have assumed $\alpha^2 \gg \beta$.
It can be shown then that
$\mu_{cr}^{ec}=\frac{\beta^2}{6 \sqrt{3}} \alpha \ll 1$ and 
$\mu_{cr}^{em} \rightarrow 1$ so that
the single 
root solution is applicable for almost all pitch angles with the root
\beq\label{kres} 
k_j\simeq -\beta^{-\frac{1}{3}}\alpha^{\frac{2}{3}}\mu^{-\frac{1}{3}}, \ 
\ \ \ 1 > \mu > \mu_{cr}^{ec}.
\eeq
In the limit of $\mu \rightarrow 0$ equation (\ref{kres}) has
three roots one of which is very large ($k \rightarrow \infty$)
compare to the other two roots which are
\beq\label{k2}
k_j \simeq \pm\frac{\sqrt{2}\alpha}{\beta},\ \ \ \mu < \mu_{cr}^{ec}.
\eeq
This expression is nearly identical to the value of $k_j$ obtained
from equation (\ref{kres}) at $\mu = \mu_{cr}^{ec}$, so that joining this two
relations at $\mu = \mu_{cr}$ will give an approximate description for
all $\mu$.

Now substitution of these values for $k_j$ in 
equation (\ref{res}) gives the values of the resonant frequencies as
\begin{eqnarray}\label{omega}\makeatletter 
  \omega_j -1\simeq \left\{
    \begin{array}[c]{lcr}
      (\alpha\mu\beta)^{\frac{2}{3}} +
      o(\beta^2),& \mu_{cr}^{ec} < \mu < 1, & \\
       o(\beta^2),&  \mu < \mu_{cr}^{ec}.&
    \end{array}\right. 
\end{eqnarray}
Then to the first order the phase velocity of the wave is obtained
from equations~(\ref{kres}) to (\ref{omega}) to be
$\beta_{ph}=\frac{\omega_j}{k_j}\simeq\frac{1}{k_j}$. Combining this
with the dispersion relation (\ref{disp}) we obtain the group
velocity
\begin{eqnarray}\label{grvel}\makeatletter 
  \beta_{gr}=\frac{d\omega}{dk} \simeq \frac{2 \alpha^2}{k^3} \simeq \left\{
    \begin{array}[c]{lcr}
      2 \mu\beta +
      o(\beta^2),&\mu_{cr}^{ec} < \mu < 1, & \\
      \frac{\beta^3}{\sqrt{2}\alpha},&
      \mu < \mu_{cr}^{ec}. & 
    \end{array}\right. 
\end{eqnarray}

Using the above expressions we can now evaluate the Fokker-Planck
coefficients and their ratios in the non-relativistic
limit. For the single root case these ratios are simplified to
$R_1=R_2^2=\omega^2/(\beta k-\mu\omega)^2$ and in the small range
of $\mu<\mu^{ec}_{cr}$ we can use the asymptotic relation
of equation (\ref{rzeromu}).
Substituting of equations (\ref{kres}) through (\ref{grvel}) into
equation (\ref{ratio}) gives the non-relativistic values of these ratios as
\beq\label{asrz}
R_1 \simeq\ \left\{
\begin{array}[c]{lcr}
  \frac{1}{2 \alpha^2}, 
  \ \ \ &\mu < \mu_{cr}^{ec} & \\ \nonumber
  \frac{\mu^{\frac{2}{3}}}
  {(\alpha^{\frac{2}{3}} \beta^{\frac{2}{3}} - \mu^{\frac{4}{3}})^2},
  \ \ \ &\mu_{ce}^{ec} <\ \mu < 1.&
\end{array}\right.
\eeq

These values agree well with the numerical results shown on Figures
\ref{ratmu} to \ref{ratmu08} at low energies.

\subsection{Acceleration Time}

\begin{figure}[tbp]
  \psfig{file=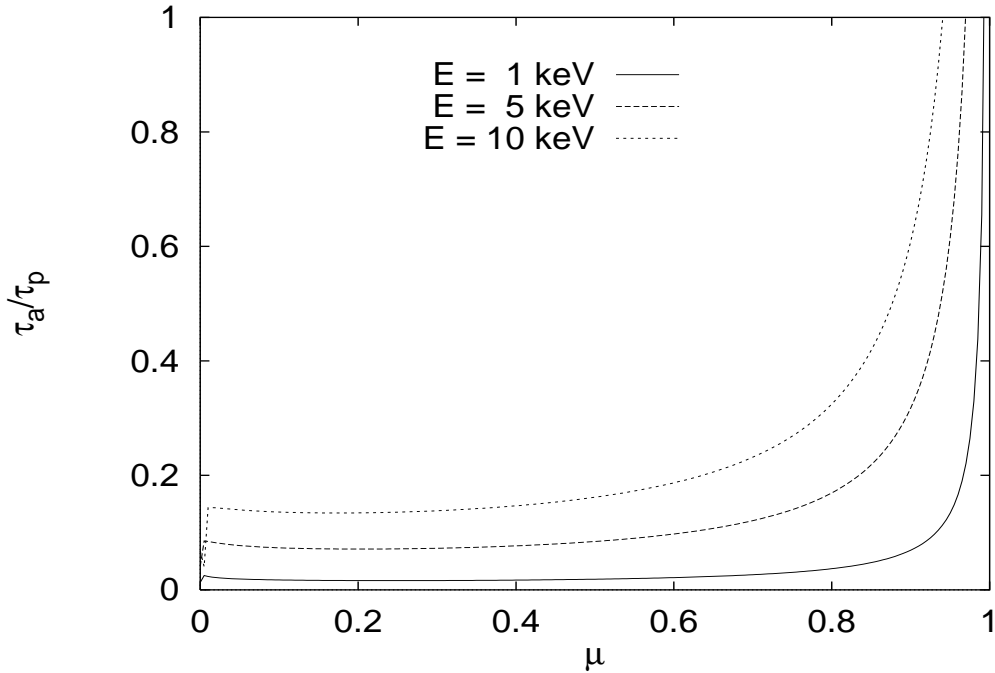,width=5.5in,height=3.5in,angle=0}
\caption{
  Dependence of acceleration time $\tau_a = (p^2/D_{pp})$ in units of
  $\tau_p$ on pitch angle $\mu$ in plasma with $\alpha = 1$ and $q =
  5/3$ for electrons with different kinetic energies. }
\label{tacm}
\end{figure}

\begin{figure}[tbp]
  \psfig{file=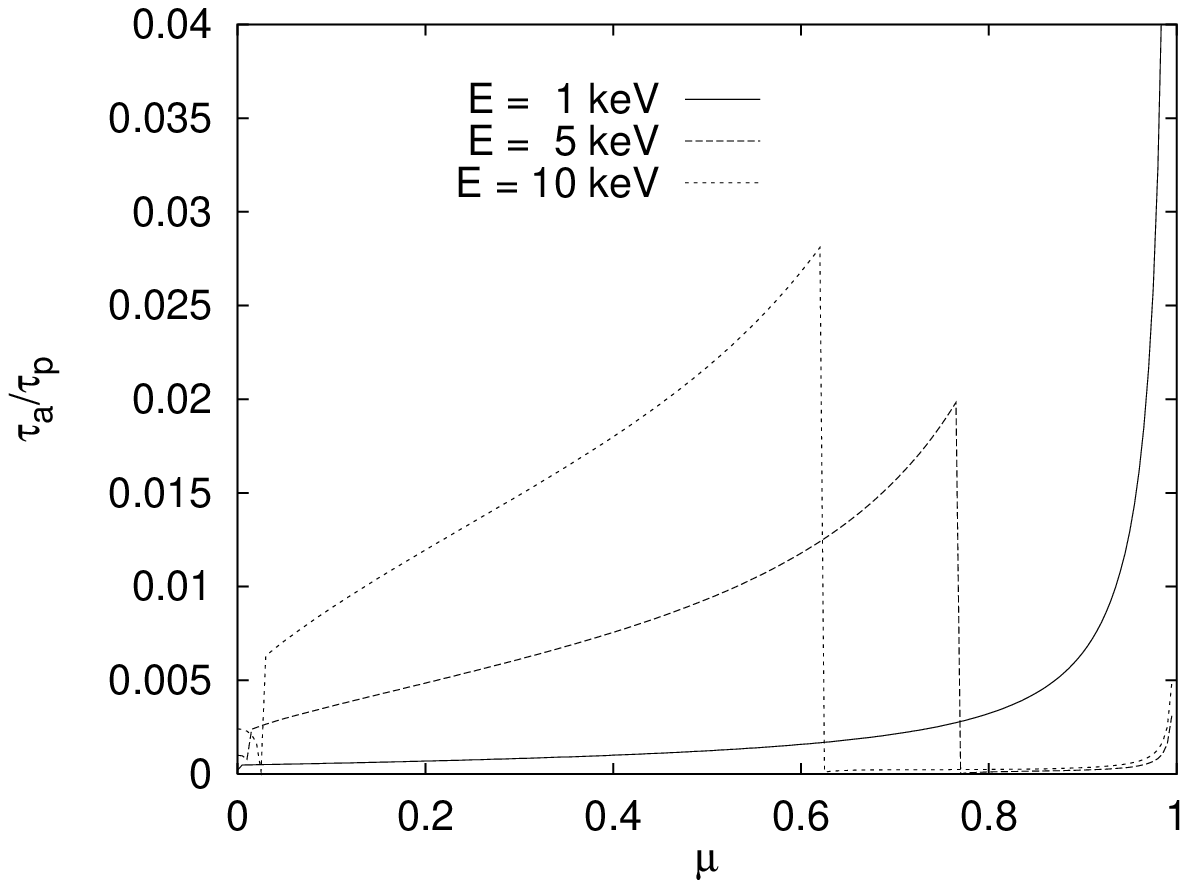,width=5.5in,height=3.5in,angle=0}
\caption{
  Same as Figure \protect\ref{tacm} with $\alpha = 0.2$ and $q =
  5/3$.}
\label{lowa}
\end{figure}

For non-relativistic electrons and especially for $\alpha < 1$ the
ratios $R_1$ and $R_2$ exceed unity and the acceleration process is 
described by
equation (\ref{newtr}). We can therefore define an acceleration time
$\tau_a = p^2/D_{pp}$. [Note that this is different from $\tau_{ac} =
1/\kappa_3$ defined for relativistic regime in combination with 
equation (\ref{trans}); see below].
Figure \ref{tacm} and \ref{lowa} show the dependence of this time 
(in units of $\tau_p$) on $\mu$
at three different
values of kinetic energy in a plasma with $q = 5/3$ and 
$\alpha =1$ and $\alpha =0.2$, respectively.  
For $\alpha \geq 1$ the critical pitch angle $\mu_{cr}^{em} = 1$ 
and the electrons
interact mainly with the whistler and electron-cyclotron waves and not with
the electromagnetic branch.
Over the wide range of $\mu$ the resonant 
interaction occurs only with backward 
(for $\mu >0$) or forward ($\mu < 0$) moving whistler waves.
For $\alpha =0.2$ we have $\mu_{cr}^{em} < 1$ (Figure \ref{lowa}) 
and there are three distinct regions on the plot with two 
discontinuous changes.
Note that the acceleration time of
electrons with pitch angles $\mu \geq \mu_{cr}^{em}$ becomes large for all
energies because of the dependence $D_{pp} \propto (1-\mu^2)$.

Using values of resonant wave vectors obtained 
in equation (\ref{dmm})
we get the following expressions for the
acceleration time
\begin{eqnarray}\label{acct}\makeatletter 
  \frac{\tau_a(\mu)}{\tau_p} \simeq \left\{
    \begin{array}[c]{lcr}
      \frac{\mu^{(1-q)/3}}{1-\mu^2} \alpha^{\frac{2(q+2)}{3}}
      \beta^{\frac{7-q}{3}},& \mu_{ce}^{ec} \ll \mu < \mu_{cr}^{em}
      \approx 1 & \\
      2^{\frac{q+1}{2}} \alpha^{(q+1)}\ \beta^{3 - q},& 
      \mu < \mu_{cr}^{ec}&
    \end{array}\right. 
  \def\@eqnnum{\relax}
\end{eqnarray}

\begin{figure}[tbp]
  \psfig{file=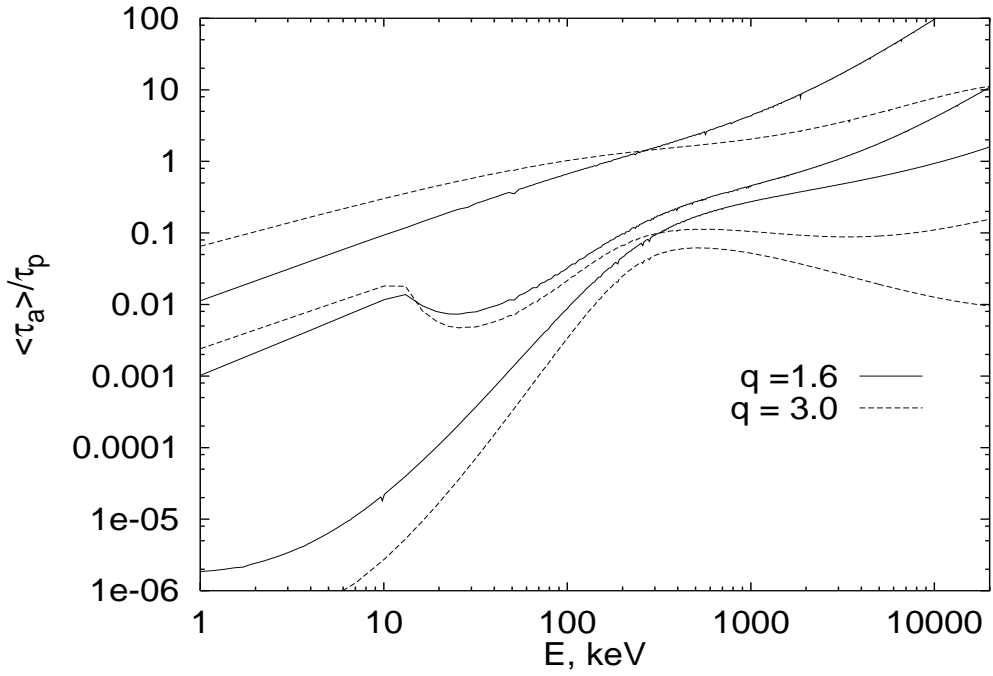,width=5.5in,height=3.5in,angle=0}
\caption{
Dependence of averaged acceleration time defined for low-
energetic electrons 
in units of $\tau_p$ (equation \protect\ref{avt}) 
on kinetic energy of electron in plasmas with two different values of 
spectral index $q$ and
$\alpha = $ 1, 0.3 and 0.1 from top to bottom.}
\label{avta}
\end{figure}

To demonstrate the dependence of the acceleration time on electron
energy alone we plot on Figure \ref{avta} the averaged acceleration time 
\beq\label{avt} 
\langle\tau_a\rangle = \frac{2p^2}{\int_{-1}^{1} d \mu D_{pp}(\mu)}
\eeq
From equations (\ref{acct}) we also obtain an
approximate analytic expression 
\beq\label{anavt}
\frac{\langle\tau_a\rangle}{\tau_p} \simeq \frac{(2+q)(8+q)}{18} \alpha^{\frac{2(q+2)}{3}} \beta^{\frac{7-q}{3}},
\eeq
where we have ignored the small contribution from electrons 
with $\mu > \mu_{cr}^{em}$
because in most cases $\mu_{cr}^{em}$ is close to $1$.
This approximation agrees with 10\% precision with numerical results
shown on Figure \ref{avta} for values of $\alpha \geq 0.3$ and
electron energies of order of keV.

Simple analytic solutions are not possible 
for the plasma conditions when a significant
fraction of electrons is able to interact with the electromagnetic branch.
As can be seen from Figure \ref{avta} in the case of small $\alpha$ the
averaged acceleration time is a complicated function of
energy and plasma parameters. In order to determine the highest energy
for which the approximate analytic expression (\ref{anavt}) is valid
one should refer to Figure \ref{goft}. For example, for 
$\alpha = 0.3$ we obtain $E_{cr} \simeq 10$ keV. This is in agreement with
Figure \ref{avta} which shows the correct power law dependence
below 10 keV.

\section{RELATIVISTIC APPROXIMATIONS}

To complete our investigation of the Fokker-Planck equation
(\ref{f-p}) in application to the electrons in a turbulent plasma in
this section we consider the case of the relativistic electrons for
which $R_{1,2}(\mu, p) \ll 1$ and the transport equation (\ref{trans})
is applicable. For this regime the equations (\ref{kappa1}) and
(\ref{kappa3}) define the average
scattering time $\tau_{sc}$ and, now the longer acceleration time $\tau_{a}$
as
\beq\label{sctime}
\tau_{sc}= \int_{-1}^{1} d\mu (1-\mu^2)^2/D_{\mu\mu},
\eeq
\beq\label{actime}
\tau_{ac}= \frac{2}{\int_{-1}^{+1}d\mu D_{\mu\mu} (R_1-R_2^2)}.
\eeq

\begin{figure}[tbp]
  \psfig{file=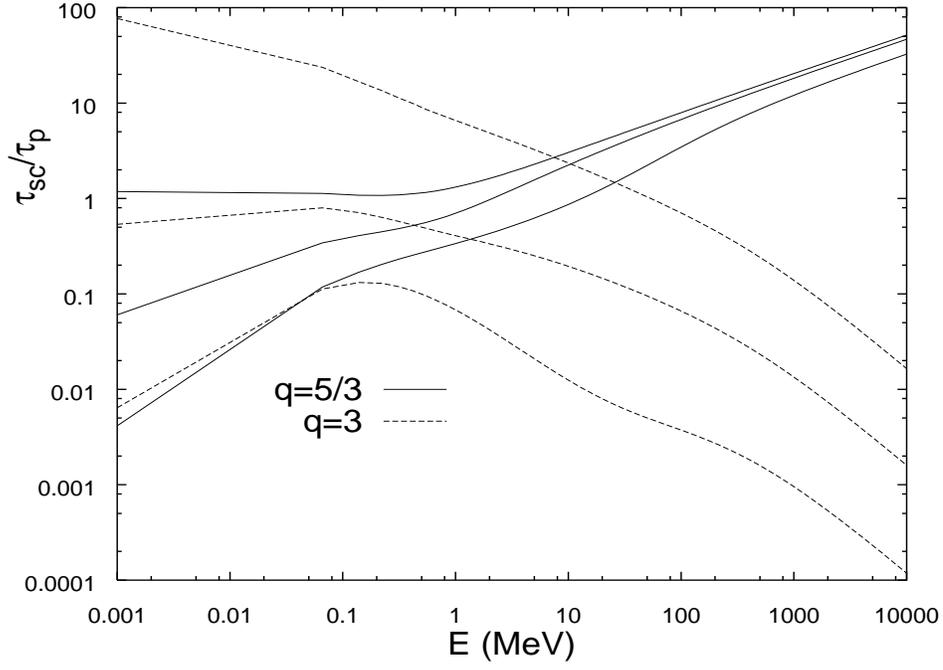,width=5.5in,height=3.5in,angle=0}
\caption{
Same as Figure \protect\ref{avta} except for 
$\tau_{sc}$ from equation (\protect\ref{sctime})
and for $\alpha = $ 10, 1.0 and 0.1 from top to bottom.}
\label{scat}
\end{figure}

\begin{figure}[tbp]
  \psfig{file=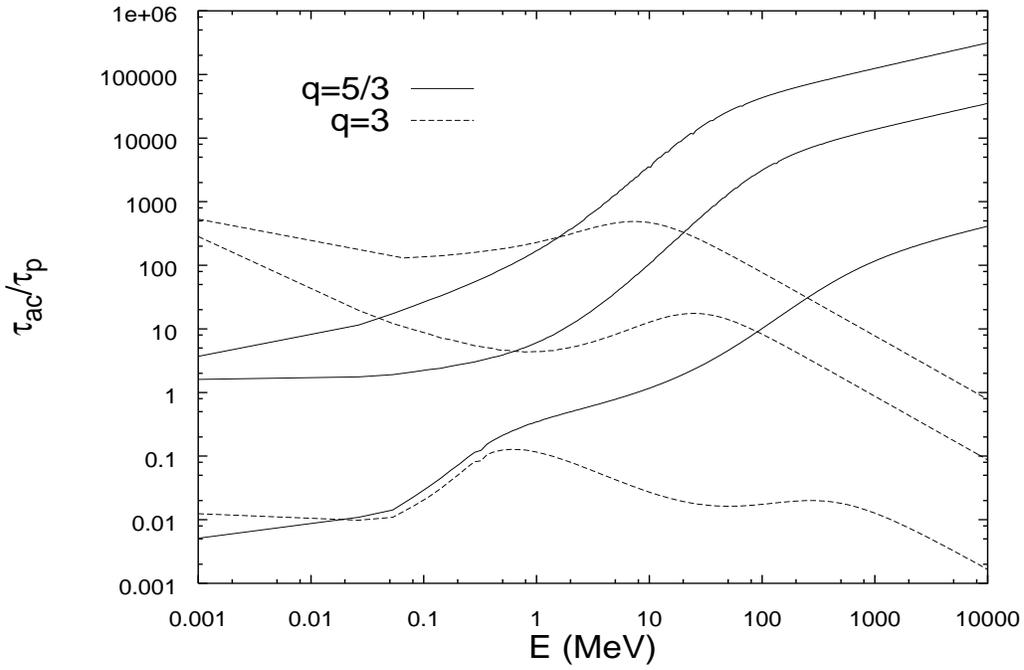,width=5.5in,height=3.5in,angle=0}
\caption{
Same as Figure \protect\ref{avta} except for 
$\tau_{ac}$ from equation (\protect\ref{actime})
and for $\alpha = $ 10, 1.0 and 0.1 from top to bottom.}
\label{relact}
\end{figure}

Figures \ref{scat} and Figure \ref{relact} show the variation with energy
of these times (in units of $\tau_p$)
obtained from numerical integration of the above equations
for different values of the plasma parameter $\alpha$ and spectral index
$q$ of the plasma turbulence. The scattering time 
is similar to the scattering time $\tau_{sc} =
1/\langle D_{\mu \mu} \rangle$ one would define in the non-relativistic
regime. Thus the curves in Figure \ref{scat} provide a good estimate
of scattering time scale at all energies. However, the acceleration
time $\tau_{ac}$ in equation (\ref{actime}) is different from the
low-energy definition $\tau_{ac} = p^2/D_{pp}$ used in \S 5.1.
These two definitions differ by the presence of $R_2^2$ term in the
denominator of equation (\ref{actime}). It turns out that for
relativistic electrons in plasmas with
small Alfv\'en velocity $R_2 \simeq R_1 \ll 1$ so that 
$R_2^2 \ll R_1$ and the two equations give very similar results. This can
also be seen from comparison of Figures \ref{avta} and \ref{relact}.
As we have shown in \S 4, this condition is not true in the
non-relativistic regime so that equation (\ref{actime}) when extended
to low energies is in error by several orders of magnitude. 

From figures \ref{avta}, \ref{scat} and \ref{relact} we see several
distinct regions of energy with different behavior of the
curves. 

\vspace{-3mm}
\paragraph{a) Extremely Relativistic Energies:}
$\gamma \gg \delta^{-1}\beta_a \ln{\beta_a^{-1}}$. \\
The power-law energy dependence ($\gamma^{2-q}$) of $\tau_{sc}$ and 
$\tau_{ac}$ seen in these figures at high energies is well known in
the literature (e.g. Schlickeiser
1989; Schlickeiser et al. 1991).
Relativistic electrons interact with Alfv\'en waves
which have a simple dispersion relation $\omega = \beta_a k$. Recall
that $\delta = m_e/m_i$ and $\beta_a = \sqrt{\delta}/\alpha$
is the Alfv\'en speed in units of speed of light, which we assume to
be less than one.
These electrons also interact with long-wavelength whistler waves
which, to the first order in $\beta_a$, have the same
dispersion relation as Alfv\'en waves. For both waves the 
resonant frequencies $\omega_R \ll \delta$.

Below we derive analytic expressions which show the above power-law
dependence as
well as the dependence of $\tau_{sc}$ and $\tau_{ac}$ on the plasma 
parameters $\alpha$ and $q$.
Using the above simplified dispersion relation and the resonant condition 
(\ref{res}) we  get two resonant values for the frequency
and the wave vector:
\begin{eqnarray}\label{krel}
k_1\simeq -\frac{1}{\gamma}\frac{1}{\mu+\beta_a},\ \ \ 
\omega_1 \simeq \frac{1}{\gamma}\frac{\beta_a}{\mu+\beta_a},\ \ \ \\ 
k_2\simeq -\frac{1}{\gamma}\frac{1}{\mu-\beta_a},\ \ \ 
\omega_2 \simeq -\frac{1}{\gamma}\frac{\beta_a}{\mu-\beta_a}.
\end{eqnarray}
Note that the demarcation energy for this case comes from the
condition $\omega_R \ll \delta$ and is different from the usually
assumed limit $\gamma \gg \delta^{-1}$.
Substitution of the above solutions in equations (\ref{sctime}) and 
(\ref{actime}) gives the asymptotic expressions for scattering and
acceleration times 
\beq\label{reltimes}
\frac{\tau_{sc}}{\tau_p} \simeq \frac{1}{4} \gamma^{2-q} I_{s_{sc}}(
\beta_a), \ \ \ \ 
\frac{\tau_{ac}}{\tau_p} \simeq \frac{1}{16} \gamma^{2-q} \frac{1}
{\beta_a^{2} I_{s_{ac}}(\beta_a)}, 
\eeq
where
\beq
I_s(\beta_a) =  \int_{0}^{1}\frac{(1-\mu^2)d\mu}
{|\mu+\beta_a|^{s}+|\mu-\beta_a|^{s}},
\eeq
and the power indices $s_{sc} = q-1,\ \ s_{ac} = 1-q$ for scattering 
and acceleration times, respectively. 

One can show that in the case of small $\beta_a$ the function $I_s(\beta_a)$
simplifies to:
\begin{eqnarray}\label{II}\makeatletter 
  I_{s}(\beta_a) \simeq \left\{
    \begin{array}[c]{lcr}
        (2 \beta_a)^{1-s}\int_{0}^{1}   
        \frac{(1-t)^{s-2}+(1+t)^{s-2}}{1+t^{s}} dt,& s>1 & \\
        - 1/2 \ln(\beta_a/4), & s = 1 & \\
        \frac{1}{(1-s)(3-s)} + 
       o(2\beta_a)^{1-s}, &s<1&
    \end{array}\right. 
  \def\@eqnnum{\relax}
\end{eqnarray}
Now we can use the above results in equations (\ref{reltimes}) to
obtain the dependence of relativistic times on plasma parameters.
We will consider spectral indices in the range $1 < q < 4$. 
In the case $ q > 2$ the scattering time for electrons of relativistic
energies has a power law dependence on plasma parameter, 
$\tau_{sc}\propto \alpha^{q-2}$.
In the opposite case of $q<2$ we have almost no
dependence of scattering time on $\alpha$.
We can see these behavior in Figure \ref{scat} for $q=5/3$ and $q=3$.

To the highest order in $\beta_a$ and for all $q$ in the above range,
$\tau_{ac} \propto \beta_a^{-2}$. Thus 
the relativistic acceleration time is proportional to
$\alpha^2$ and this dependence 
can be seen on Figure \ref{relact}. The above results indicate that 
the ratio $R_1 \simeq \tau_{sc}/\tau_{ac}$ has a strong dependence 
on $\alpha$
and $q$ which
leads to the fact that in low-$\alpha$ plasmas with spectral index
$q<2$ this ratio is of order of 1 even for relativistic
electrons and equation (\ref{trans}) is not valid. 

\vspace{-3mm}
\paragraph{b) Intermediate Regime:}
$1   \ll \gamma \ll \delta^{-1}\beta_a \ln{\beta_a^{-1}}$. \\
As can be seen from Figure \ref{scat} and \ref{relact} at $\gamma
\simeq \delta^{-1}\beta_a \ln{\beta_a^{-1}}$ the curves begin to 
deviate from the described
power-law because the interaction with whistler mode with the higher 
wave vectors begin to dominate. In this case $1 \gg \omega_R \gg
\delta$ and the simple expressions such as
(\ref{krel}) and (\ref{reltimes}) are not possible and one has to
consider different combinations of plasma parameters in order to get
analytic expressions for scattering and acceleration times correspondent to
that particular plasma conditions. Different cases of interaction of
electrons with whistlers have been considered in
several papers (e.g. Steinacker and Miller 1992, Hamilton and Petrosian 1992).

\vspace{-3mm}
\paragraph{c) Non-relativistic Energies:} $ (\gamma -1) \ll 1$. \\
Lastly, for non-relativistic electrons considered in \S 5 
the resonant frequency 
$\omega_R \rightarrow 1$ and electrons interact most effectively 
with short-wavelength whistlers. 
As we described in \S 3.1, in plasmas with $\alpha < 1$ electrons also
interact with lower frequency electromagnetic branch of plasma waves.
We see the
corresponding change of behavior of the curves on Figures \ref{avta} 
and \ref{scat} as we approach non-relativistic energies.
The interaction with electromagnetic waves is allowed only for 
electrons with energy greater
than the critical energy $E_{cr}$ shown on Figure \ref{goft}. 
In this regime the dependence of both scattering and acceleration
times on energy deviates significantly from a power-law.
For energies less than $E_{cr}$ equation (\ref{avt}) becomes valid and
we again can see in Figure \ref{avta} the power-law dependence of the
acceleration time on kinetic energy of electron.

\section{SUMMARY}

The importance of the stochastic acceleration of high energy charged
particles by turbulent plasma waves is well known. 
In this work we investigate the possibility of
acceleration of the low energy background (often thermal) electrons by
this process. We use the well known formalism developed over the years
and specifically the formalism proposed by Schlickeiser and DP. 
In this paper we consider interaction of electrons with plasma waves 
propagating
along the magnetic field lines. At all energies the plasma parameters
which determine the acceleration rate of the electrons are the
value of the magnetic field, the energy density of the plasma
turbulence, the spectral index $q$ of the waves and most importantly 
the plasma parameter $\alpha$ which is equal to the 
ratio of plasma frequency to gyrofrequency of electrons.
At low energies interaction with the electron cyclotron and
electromagnetic branch of plasma waves become important. The second
interaction is possible only when $\alpha < 1$ and above the
critical energies shown in Figure \ref{goft}.
The electrons which do not have enough energy to interact with the
electromagnetic waves may still be in resonance with whistler and/or 
electron-cyclotron modes.

From the above consideration we show that the ratio of the pitch angle
to momentum (or energy) diffusion rates, or alternatively, the ratio of
the acceleration to scattering times (equation \ref{r}) varies
strongly with the plasma parameters and the pitch angle and energy of
the electron. We give asymptotic analytic expression for these ratios
showing their dependence on these parameters. In particular we
show that this ratio becomes greater than one at lower energies which
indicates that the usual transport equation derived for cosmic rays is
not applicable for electrons for a wide range of energies. An approximate
"critical" value of electron energy above which the dynamics of
the electrons may be described by this
transport equation is determined as a function of plasma parameter.
We propose a new transport equation for non-relativistic electrons in
low-$\alpha$ plasmas.

We also show numerically and through asymptotic analytic expressions
that when the above ratio is greater than one the acceleration time
scale decreases with decreasing energy and could be very short
for magnetic field of 100 G and turbulent energy
density of less than $10^{-4}$ of that of magnetic field energy
density.  
To complete the discussion we also consider the diffusion rates for
relativistic electrons for which the above mentioned ratio is less 
than one. We give analytic and numerical results on the acceleration and
scattering time scales for different energies and plasma parameters.

Based on these results
we suggest the following scenario for acceleration of the background
plasma electrons. The low energy (background) electrons can be accelerated 
by whistler, high-frequency electromagnetic or electron-cyclotron waves 
without a significant change in their pitch angle
over time scale of order of the acceleration time 
$\tau_a = p^2/D_{\mu \mu}(\mu)$. This process holds till
electrons  reach an energy where the scattering time becomes comparable
to the acceleration time. Then intensive scattering along with
acceleration takes place and for this stage we have to take into
account all of the terms in the Fokker-Planck equation to describe the time
evolution of the distribution function. When electrons accelerate
to high enough energies (typically energies of tens MeV) their scattering time
becomes much less than all other time scales, the distribution function
becomes nearly isotropic and 
the well-known transport equation (\ref{trans}) becomes applicable.
For the two limiting cases
with simple transport equations analytic expressions for scattering 
and acceleration times
as a function of energy, plasma parameter and 
turbulence spectral index can be used.

In this paper we have considered interaction of electrons with plasma
waves propagating along the magnetic field lines, in a ``cold''
background plasma. In subsequent papers we will consider waves with
oblique propagation angles and include the effects of the finite
temperature of the background plasma.

We acknowledge J. Miller and P. Sturrock for useful discussions.
This work was supported by NSF grant ATM 93-11888 and NASA grant NAGW 1976.



\begin{thebibliography}{}
\bibitem{1} Achatz, U., Dr\"oge, W., Schlickeiser, R., Wibberenz, G. 1993,
J. Geophys. Res., 98, 13261
\bibitem{2}Achterberg A. 1981, Astron.Astrophys., 97,  259. 
\bibitem{3}Beaujardiere, J.-F. de la and Zweibel E.G. 1989, ApJ, 336, 1059.
\bibitem{4}Benz A. 1977, ApJ, 221, 270. 
\bibitem{5}Benz A. and Smith D. 1987, Solar Phys., 107, 299.
\bibitem{6} Dung, R. and Petrosian, V. 1994, ApJ, 421, 550
\bibitem{7} Hamilton, R.J. and Petrosian, V. 1992, ApJ, 398, 350
\bibitem{8} Jaekel, U. and Schlickeiser, R. 1992, J. Phys., 18, 1089
\bibitem{9} Jokipii, J.R. 1966, ApJ, 146, 480
\bibitem{10} Kennel, C.F. and Engelmann, F. 1966, Phys. Fluids, 9, 2377
\bibitem{11} Kennel, C.F. and Wong H.V. 1967, Plasma Phys, 1, 75
\bibitem{12} Kirk, J.G, Schlickeiser, R. and Schneider, P. 1988, ApJ, 328, 269 
\bibitem{13} Lerche, I. 1968, Phys. Fluids, 11, 1720
\bibitem{14}Melrose D.B. 1974, Solar Phys., 37, 353.
\bibitem{15}Melrose D.B. 1980, Plasma Astrophysics, (Gordon and 
Breach, New York).
\bibitem{16}Miller J.A. and Ramaty R. 1987, Solar Phys., 113, 195.
\bibitem{17}Miller J.A. 1991, ApJ. 376, 342.
\bibitem{18}Schlickeiser, R. 1989, ApJ, 336, 243
\bibitem{19}Steinacker, J. and Miller, J.A. 1992, ApJ., 393, 764
\bibitem{20}Stix, T.H. 1992, Waves in Plasma, (AIP, New York)
\bibitem{21}Sturrock, P.A. 1994, Plasma Physics, (Cambridge University Press)
\bibitem{22} Tademaru, E. 1969, ApJ, 158, 959
\end{thebibliography}
\end{document}